\documentclass[showpacs,twocolumn,preprintnumbers,amsmath,amssymb]{revtex4-1}
\usepackage{graphicx}% Include Figure  files
\usepackage{dcolumn}% Align table columns on decimal point
\usepackage{bm}% bold math
\usepackage{color}
\usepackage{amsmath}
\usepackage{upgreek}
\usepackage{jabbrv}
\usepackage{caption}
\usepackage{subcaption}
\usepackage[raggedright]{titlesec}

\begin{document}
\title{ Optical phase mining by adjustable spatial differentiator }
\author{ Tengfeng Zhu$^{1}$, Junyi Huang$^{1}$, Zhichao Ruan$^{1,2}$}\email{zhichao@zju.edu.cn}
\affiliation{$^1$ Interdisciplinary Center for Quantum Information, State Key Laboratory of Modern Optical Instrumentation, and Zhejiang Province Key Laboratory of Quantum Technology and Device, Department of Physics, Zhejiang University, Hangzhou 310027, China. \\
$^2$ College of Optical Science and Engineering, Zhejiang University, Hangzhou 310027, China \\
}
\begin{abstract}
Phase is a fundamental resource for optical imaging but cannot be directly observed with intensity measurements. The existing methods to quantify a phase distribution rely on complex devices and structures. Here we experimentally demonstrate a phase mining method based on so-called adjustable spatial differentiation, just generally by analyzing the polarization in light reflection on a single planar dielectric interface. With introducing an adjustable bias, we create a virtual light source to render the measured images with a shadow-cast effect. We further successfully recover the phase distribution of a transparent object from the virtual shadowed images. Without any dependence on resonance or material dispersion, this method directly stems from the intrinsic properties of light and can be generally extended to a board frequency range.
\end{abstract}

\maketitle

\section{introduction}

Phase distribution plays the key role in biological microscopic imaging and X-ray imaging but cannot be directly observed with intensity measurements, where the information of objects or structures are mostly stored in phase rather than intensity. Zernike invented the phase contrast microscopy \cite{zernike1955discovered} to render these transparent objects but without the ability to quantify the phase distribution. Later, in order to enhance the image contrast, Nomarski prisms were created for differential interference contrast (DIC) imaging \cite{allen1969zeiss} which offers the phase gradient information \cite{McIntyre:09, davis2000image, furhapter2005spiral, jesacher2005shadow, qiu2018spiral}. Further, various quantitative phase measurement technologies were developed for mapping the optical thickness of specimens \cite{Ikeda:05, Popescu:06, Zheng:17, park2018quantitative}. These methods usually rely on complex modulation devices in the spatial or spatial frequency domains, resulting in the difficulties of optical alignment and adjustment.

In recent years, optical analog computing of spatial differentiation has attracted great attention, which enables an entire image processes on a single shot  \cite{Silva2014performing, PorsNielsenBozhevolnyi14, doskolovich2014spatial, ruan2015spatial, Abdollah15, Chizari2016, Youssefi:16, HwangDavis16, Fang2017On, Wu:17, hwang2018plasmonic}. Various optical analog computing devices were designed for edge detection \cite{zhu2017plasmonic, Fang:18, saba2018two, dong2018optical, Roberts:18, guo2018photonic, PhysRevLett.121.173004, Zhu2019Generalized, Zhou11137, Guo:18, PhysRevApplied.11.064042, PhysRevLett.123.013901}. In particular, Silva \textit{et al.} theoretically proposed realization of optical mathematical operations with metamaterials, as well as the Green's function slabs \cite{Silva2014performing}. Later a subwavelength-scale plasmonic differentiator of a 50-nm-thick silver layer was experimentally demonstrated and applied for real-time edge detection \cite{zhu2017plasmonic}. Most recently, spatial differentiation methods were developed based on the geometric phases, including the Rytov-Vladimirskii-Berry phase \cite{Zhu2019Generalized} and the Pancharatnam-Berry phase \cite{Aleman-Castaneda:19, Zhou11137}. However, so far the current application of optical spatial differentiation is only limited to edge detection. As the fundamental difficulty to determine the phase of electric fields by measuring their intensities, one can neither determine the phase distributions from the edge-enhanced signal.

In this article, we propose an adjustable spatial differentiation to characterize and quantitatively recover the phase distribution, just by simply analyzing the polarization in light reflection on a single planar dielectric interface. By investigating the field transformation between two linear cross polarizations, we experimentally demonstrate that the light reflection enables an optical analog computing of spatial differentiation with adjustable direction. We show that this effect is related to the angular Goos-H{\"a}nchen (GH) shift \cite{ra1973reflection, chan1985angular, merano2009observing} and the Imbert-Fedorov (IF) shift \cite{fedorov1955k, PhysRevD.5.787}. Furthermore, by tuning a uniform constant background as the bias, we create a virtual light source to render the measured images with a shadow-cast effect and further quantify  the phase distribution of a coherent field. Without complex modulation devices \cite{ Ikeda:05, Popescu:06, McIntyre:09, Zheng:17, davis2000image, furhapter2005spiral, jesacher2005shadow, qiu2018spiral}, our method offers great simplicity and flexibility. Importantly, since the proposed method is irrelevant to resonance or material dispersion, it works in general wavelength with large temporal bandwidths which is very suitable for high-throughput real-time image processing.

\begin{figure}
\centerline{\includegraphics[width=3.2in]{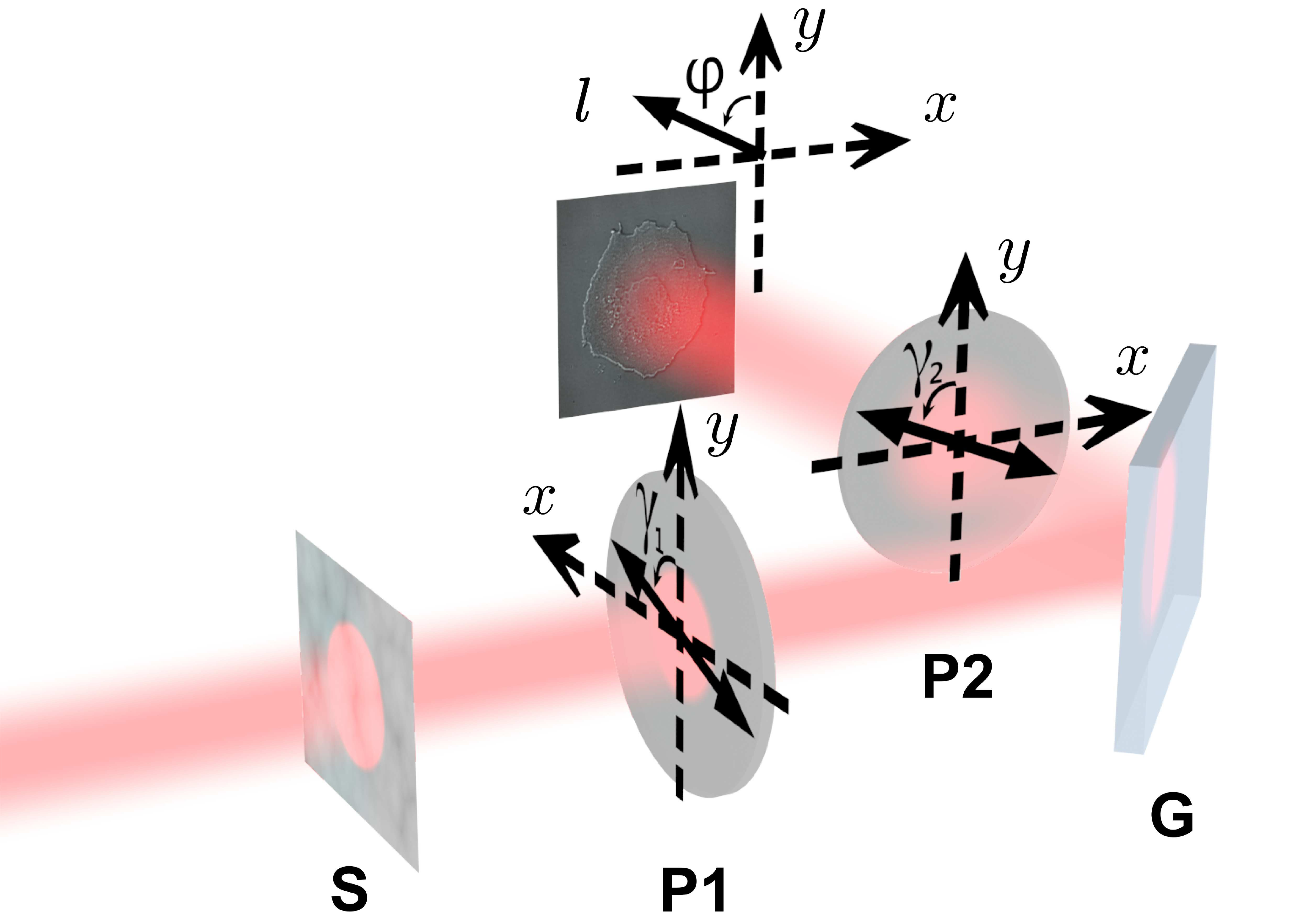}}%Here is how to import EPS art
\caption{\label{fig:1} Schematic of the phase-mining method based on polarization analysis in light reflection on a dielectric interface, e.g. an air-glass interface. A phase object S is uniformly illuminated and then the light is polarized by a polarizer P1 and reflected on the surface of a glass slab G. By analyzing the polarization of the reflected light with a polarizer P2, the differential contrast image of S appears at the imaging plane of the system and can exhibit a shadow-cast effect. The polarizers P1 and P2 are orientated at the angles ${\gamma _1}$  and ${\gamma _2}$. The direction of the spatial differentiation is along  $l$ and indicated with an angle $\varphi $.}
\end{figure}

\section{Results}
\subsection{Adjustable spatial differentiation}

Schematically shown as Fig.~\ref{fig:1}, the proposed phase mining scheme only includes a dielectric interface as a planar reflector and two polarizers P1 and P2, where the incident light is ${S_{in}} = {e^{ - i\psi \left( {x,y} \right)}}$ assuming with a phase distribution $\psi \left( {x,y} \right)$. Next we show that such a light reflection process can transform the original uniform intensity image with invisible phase structure into a structured contrast image, by optical computing of spatial differentiation to the incident electric field. Moreover, the direction of the spatial differentiation can be further adjusted with different combinations of the polarizer orientation angles.

To specifically depict the transformation, we decompose the incident (output) beam into a series of plane waves as ${S_{in\left( {out} \right)}} = \iint {{{\tilde S}_{in\left( {out} \right)}}}({k_x},{k_y})\exp (i{k_x}x)\exp (i{k_y}y)d{k_x}d{k_y}$ according to spatial Fourier transform. Then the spatial transformation between the measured field and the original one is determined by a spatial spectral transfer function $H \equiv {{{{\tilde S}_{out}}\left( {{k_x},{k_y}} \right)} \mathord{\left/
 {\vphantom {{{{\tilde S}_{out}}\left( {{k_x},{k_y}} \right)} {{{\tilde S}_{in}}\left( {{k_x},{k_y}} \right)}}} \right.
 \kern-\nulldelimiterspace} {{{\tilde S}_{in}}\left( {{k_x},{k_y}} \right)}}$. We denote the orientation angles of the polarizers P1 and P2 as  ${\gamma _1}$ and ${\gamma _2}$, respectively. The spatial spectral transfer function can be written as (see Supplementary Section 1)
\begin{equation}
\begin{aligned}
H &= {r_p}\sin {\gamma _1}\sin {\gamma _2} + {r_s}\cos {\gamma _1}\cos {\gamma _2}\\
 &+ \frac{{\delta \left( {{r_p} + {r_s}} \right)\sin \left( {{\gamma _2} - {\gamma _1}} \right)}}{2}{k_y},
\label{eq:1}
\end{aligned}
\end{equation}
where ${r_p}$ and ${r_s}$ are the Fresnel reflection coefficients of the p- and s-polarized plane waves, respectively. In Eq.~(\ref{eq:1}), the third term is induced by the opposite $y$-directional displacements  $\delta  = {{2\cot {\theta _0}} \mathord{\left/
 {\vphantom {{2\cot {\theta _0}} {{k_0}}}} \right.
 \kern-\nulldelimiterspace} {{k_0}}}$ for left- and right-handed circularly polarized beams, which is known as the IF shift \cite{Zhu2019Generalized, hosten2008observation}.
By considering the GH effect, we expand the  ${r_a}$ ($a = p,s$) around the incident angle ${\theta _0}$  as ${r_a} = {r_{a0}} + \frac{{\partial {r_a}}}{{\partial \theta }}\frac{{{k_x}}}{{{k_0}}}$, where ${r_{a0}}$ is the Fresnel reflection coefficient for the central plane wave, and ${k_0}$ is the wavevector number in vacuum. We note that here such a spatial dispersion during partial reflection on the dielectric interface purely leads to an angular GH shift \cite{ra1973reflection, chan1985angular, merano2009observing}.

 For a partial reflection process on a dielectric interface, we can control the orientation angle of the polarizers P1 and P2 to satisfy the cross-polarization condition
\begin{equation}
\begin{aligned}
{r_{p0}}\sin {\gamma _1}\sin {\gamma _2} =  - {r_{s0}}\cos {\gamma _1}\cos {\gamma _2}.
\label{eq:2}
\end{aligned}
\end{equation}
In such a condition, the P2 polarizer is oriented orthogonally to the polarization of the reflected central plane wave, which is first polarized by P1 and then reflected by the dielectric interface with an incident angle ${\theta _0}$. Under the cross-polarization condition, the spatial spectral transfer function becomes
\begin{equation}
\begin{aligned}
H =  - \left( {{C_1}{k_x} + {C_2}{k_y}} \right),
\label{eq:3}
\end{aligned}
\end{equation}
where ${C_1}$ and ${C_2}$ are two coefficients as ${C_1} =  - {{\left( {\sin {\gamma _1}\sin {\gamma _2}{{\partial {r_p}} \mathord{\left/
 {\vphantom {{\partial {r_p}} {\partial \theta }}} \right.
 \kern-\nulldelimiterspace} {\partial \theta }} + \cos {\gamma _1}\cos {\gamma _2}{{\partial {r_s}} \mathord{\left/
 {\vphantom {{\partial {r_s}} {\partial \theta }}} \right.
 \kern-\nulldelimiterspace} {\partial \theta }}} \right)} \mathord{\left/
 {\vphantom {{\left( {\sin {\gamma _1}\sin {\gamma _2}{{\partial {r_p}} \mathord{\left/
 {\vphantom {{\partial {r_p}} {\partial \theta }}} \right.
 \kern-\nulldelimiterspace} {\partial \theta }} + \cos {\gamma _1}\cos {\gamma _2}{{\partial {r_s}} \mathord{\left/
 {\vphantom {{\partial {r_s}} {\partial \theta }}} \right.
 \kern-\nulldelimiterspace} {\partial \theta }}} \right)} {{k_0}}}} \right.
 \kern-\nulldelimiterspace} {{k_0}}}$ and ${C_2} = {{\delta \left( {{r_{p0}} + {r_{s0}}} \right)\sin \left( {{\gamma _1} - {\gamma _2}} \right)} \mathord{\left/
 {\vphantom {{\delta \left( {{r_{p0}} + {r_{s0}}} \right)\sin \left( {{\gamma _1} - {\gamma _2}} \right)} 2}} \right.
 \kern-\nulldelimiterspace} 2}$. Equation~(\ref{eq:3}) shows that the spatial spectral transfer function is linearly dependent on the spatial frequencies ${k_x}$  and ${k_y}$, which corresponds to the computation of a directional differentiation in the spatial domain: ${S_{out}}\left( {x,y} \right) = i\left( {{C_1}\frac{{\partial {S_{in}}}}{{\partial x}} + {C_2}\frac{{\partial {S_{in}}}}{{\partial y}}} \right)$. We note that the directional spatial differentiation occurs in every oblique partial reflection case on a dielectric interface, but hardly in total internal reflection or metallic reflection cases where the complex ${r_{p0}}$  and ${r_{s0}}$ prevent the satisfaction of Eq.~(\ref{eq:2}). Besides, since ${C_1}$  and ${C_2}$  in Eq.~(\ref{eq:3}) are also complex, the directional differentiation no longer exists unless  ${{{C_1}} \mathord{\left/
 {\vphantom {{{C_1}} {{C_2}}}} \right.
 \kern-\nulldelimiterspace} {{C_2}}}$  is exactly purely real.

 Considering the original field ${S_{in}} = {e^{ - i\psi \left( {x,y} \right)}}$, the output field ${S_{out}}$  is
\begin{equation}
\begin{aligned}
{S_{out}}\left( {x,y} \right) = i\left| {\vec l} \right|\frac{{\partial {S_{in}}}}{{\partial l}} = A{e^{ - i\psi }}\frac{{\partial \psi }}{{\partial l}},
\label{eq:4}
\end{aligned}
\end{equation}
which offers a differential contrast imaging of the phase object along direction $\vec l$  defined as $\vec l = \left( {{C_1},{C_2}} \right)$. Besides, the coefficient  $A = \left| {\vec l} \right|$ in Eq.~(\ref{eq:4}) varies with different directions  $\vec l$ (see Supplementary Section 2), which is a quantitatively important coefficient when utilizing spatial differentiation along different directions in the meantime.

We note that the direction $\vec l$, described by an angle  $\varphi $ as shown in Fig.~\ref{fig:1}, is continuously adjustable with different values of ${C_1}$  and  ${C_2}$. Under a certain incident angle, the direction angle $\varphi $  varies with different pairs of  ${\gamma _1}$ and ${\gamma _2}$  that satisfy the cross-polarization condition. We note that only when the incident angle is smaller than Brewster angle, can $\varphi $ cover a complete range from $0^\circ $ to $180^\circ $. However, when the incident angle is larger than Brewster angle, ${r_{p0}}$  and ${r_{s0}}$  become both negative and hence the coefficients  ${C_1}$ and ${C_2}$  are always negative and in a limited range, so that the corresponding values of $\varphi $ are also limited.

\subsection{Bias introduction and phase mining}

We note that even though the edges of phase objects can be detected through the directional differentiation along direction $\vec l$, the sign of the differentiation ${{\partial \psi } \mathord{\left/
 {\vphantom {{\partial \psi } {\partial l}}} \right.
 \kern-\nulldelimiterspace} {\partial l}}$  cannot be distinguished, since the measured intensity is proportional to  ${I_{out}} \equiv {\left| {{S_{out}}} \right|^2} = {\left| {A{{\partial \psi } \mathord{\left/
 {\vphantom {{\partial \psi } {\partial l}}} \right.
 \kern-\nulldelimiterspace} {\partial l}}} \right|^2}$. It means the edges cannot be determined as the ridges or the troughs in the phase structure. In order to determine the sign of the differentiation, we add a uniform constant background as a bias into the proposed spatial differentiation and generate contrast images with shadow-cast effect. We show that the bias can be easily introduced and adjusted in the proposed phase quantifying scheme, in comparison to the other methods, such as traditional DIC microscopy \cite{allen1969zeiss, shaked2012biomedical} or spiral phase contrast microscopy.

 The uniform constant background is introduced by breaking the spatial differentiation requirement of Eq.~(\ref{eq:2}), that is, by rotating the polarizers for a small angle deviating from the cross-polarization condition. In this case, Eq.~(\ref{eq:2}) is changed to
\begin{equation}
\begin{aligned}
{r_{p0}}\sin {\gamma _1}\sin {\gamma _2} =  - {r_{s0}}\cos {\gamma _1}\cos {\gamma _2} + b,
\label{eq:5}
\end{aligned}
\end{equation}
where $b$ is a constant and continuously adjustable with  ${\gamma _1}$ and ${\gamma _2}$.

Following in the same way as Eqs.~(\ref{eq:3}) and ~(\ref{eq:4}), after introducing the uniform constant background, it shows that the measured reflected image is changed to   ${I_{out}} = {\left| {A{{\partial \psi } \mathord{\left/
 {\vphantom {{\partial \psi } {\partial l}}} \right.
 \kern-\nulldelimiterspace} {\partial l}} + b} \right|^2}$ (see the details in Supplementary Section 3) and becomes shadow-cast. The sign of  ${{\partial \psi } \mathord{\left/
 {\vphantom {{\partial \psi } {\partial l}}} \right.
 \kern-\nulldelimiterspace} {\partial l}}$ can be determined from the change of intensity. For example, with a positive bias  $b$, the intensity appears brighter (darker) in the areas with positive (negative) directional derivative  ${{\partial \psi } \mathord{\left/
 {\vphantom {{\partial \psi } {\partial l}}} \right.
 \kern-\nulldelimiterspace} {\partial l}}$. The biased image exhibits a shadow-cast effect, which appears like being illuminated by a virtual light source as schematically shown in Fig.~\ref{fig:4}(c).

Finally, we can recover the phase distribution  $\psi \left( {x,y} \right)$ with both the determined sign and the absolute value of ${{\partial \psi } \mathord{\left/
 {\vphantom {{\partial \psi } {\partial l}}} \right.
 \kern-\nulldelimiterspace} {\partial l}}$  . We implement the retrieval of phase distribution through a 2D Fourier method \cite{arnison2004linear}. It first requires two orthogonal directional derivatives, for example the partial derivatives ${{\partial \psi } \mathord{\left/
 {\vphantom {{\partial \psi } {\partial y}}} \right.
 \kern-\nulldelimiterspace} {\partial y}}$  and ${{\partial \psi } \mathord{\left/
 {\vphantom {{\partial \psi } {\partial x}}} \right.
 \kern-\nulldelimiterspace} {\partial x}}$, and then combines them as a complex distribution  $g\left( {x,y} \right) = {{\partial \psi } \mathord{\left/
 {\vphantom {{\partial \psi } {\partial x}}} \right.
 \kern-\nulldelimiterspace} {\partial x}} + i{{\partial \psi } \mathord{\left/
 {\vphantom {{\partial \psi } {\partial y}}} \right.
 \kern-\nulldelimiterspace} {\partial y}}$. In spatial frequency domain, its 2D Fourier transform is  $F\left[ {g\left( {x,y} \right)} \right] = i\left( {{k_x} + i{k_y}} \right)F\left[ {\psi \left( {x,y} \right)} \right]$. Therefore, the phase distribution $\psi \left( {x,y} \right)$  can be recovered as
 \begin{equation}
\begin{aligned}
\psi \left( {x,y} \right) = {F^{ - 1}}\left[ {\frac{{F\left[ {g\left( {x,y} \right)} \right]}}{{i\left( {{k_x} + i{k_y}} \right)}}} \right].
\label{eq:6}
\end{aligned}
\end{equation}

\subsection{Experimental demonstration}

\begin{figure}[t]
\centerline{\includegraphics[width=3.2in]{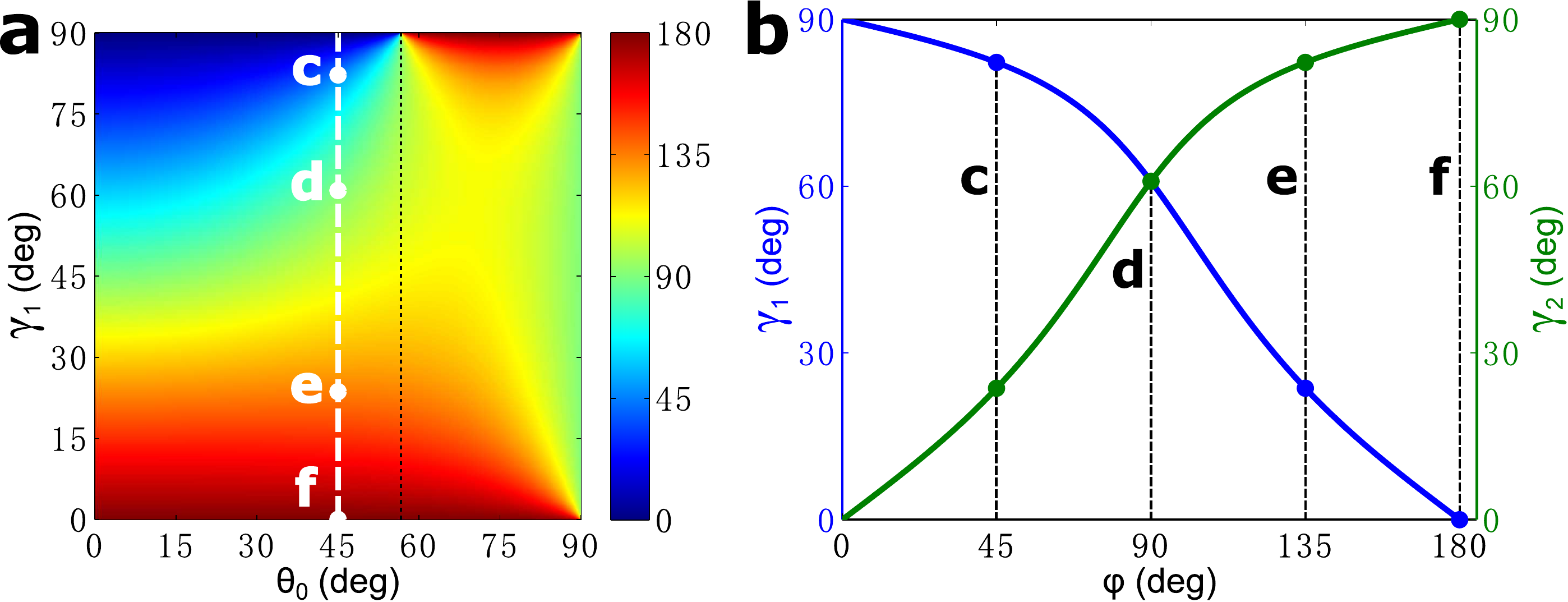}}%Here is how to import EPS art
\caption{\label{fig:2} Adjustability demonstration of the direction angle $\varphi $. As an example, here the source is a green laser with wavelength  $\lambda  = 532{\textrm{nm}}$ in vacuum and a BK7 glass slab with refractive index 1.5195 is used for reflection. (a) Adjustable range of direction angle $\varphi$  under a certain incident angle ${\theta _0}$  with different pairs of ${\gamma _1}$  and  ${\gamma _2}$ satisfying the cross-polarization condition. The white and black dashed lines correspond to ${\theta _0} = 45^\circ $  and the Brewster angle, respectively. The points (c-f) on the white dashed line correspond to  $\varphi  = 45^\circ $,  $90^\circ $, $135^\circ $  and  $180^\circ $, respectively. on the white dashed line in (a), where  ${\theta _0} = 45^\circ $. (b) Specific values of ${\gamma _1}$  and  ${\gamma _2}$ for a certain direction angle $\varphi $. The points c-f are the same as those in (a), corresponding to $\varphi  = 45^\circ $,  $90^\circ $, $135^\circ $  and  $180^\circ $, respectively.}
\end{figure}

To demonstrate the proposed phase quantifying, we use a green laser source with wavelength $\lambda  = 532{\textrm{nm}}$  and a BK7 glass slab with refractive index 1.5195 for reflection. According to Eqs.~(\ref{eq:1})-(\ref{eq:3}), we first simulate the adjustable range of  $\varphi $ under every different incident angle and the chosen orientation angle of P1. As shown in Fig.~\ref{fig:2}(a), indeed the direction angle $\varphi $  can be fully adjustable from 0 to 180 degree only when the incident angle is smaller than the Brewster angle (black dashed line).

In order to show a complete adjustable range from 0 to 180 degree, we select an incident angle ${\theta _0} = 45^\circ $  [the white dashed line in Fig.~\ref{fig:2}(a)]. We next experimentally demonstrate spatial differentiation for the direction angles  $\varphi  = 45^\circ $,  $90^\circ $, $135^\circ $  and  $180^\circ $, by choosing the appropriate orientation angles ${\gamma _1}$  and ${\gamma _2}$ (see specific values in Supplementary Section 4). By the requirement of Eq.~(\ref{eq:2}), Fig.~\ref{fig:2}(b) gives the specific values of  ${\gamma _1}$  and ${\gamma _2}$. The points (c-f) in Figs.~\ref{fig:2}(a) and (b) correspond to the cases with these four direction angles  $\varphi $, respectively.

\begin{figure}[t]
\centerline{\includegraphics[width=3.2in]{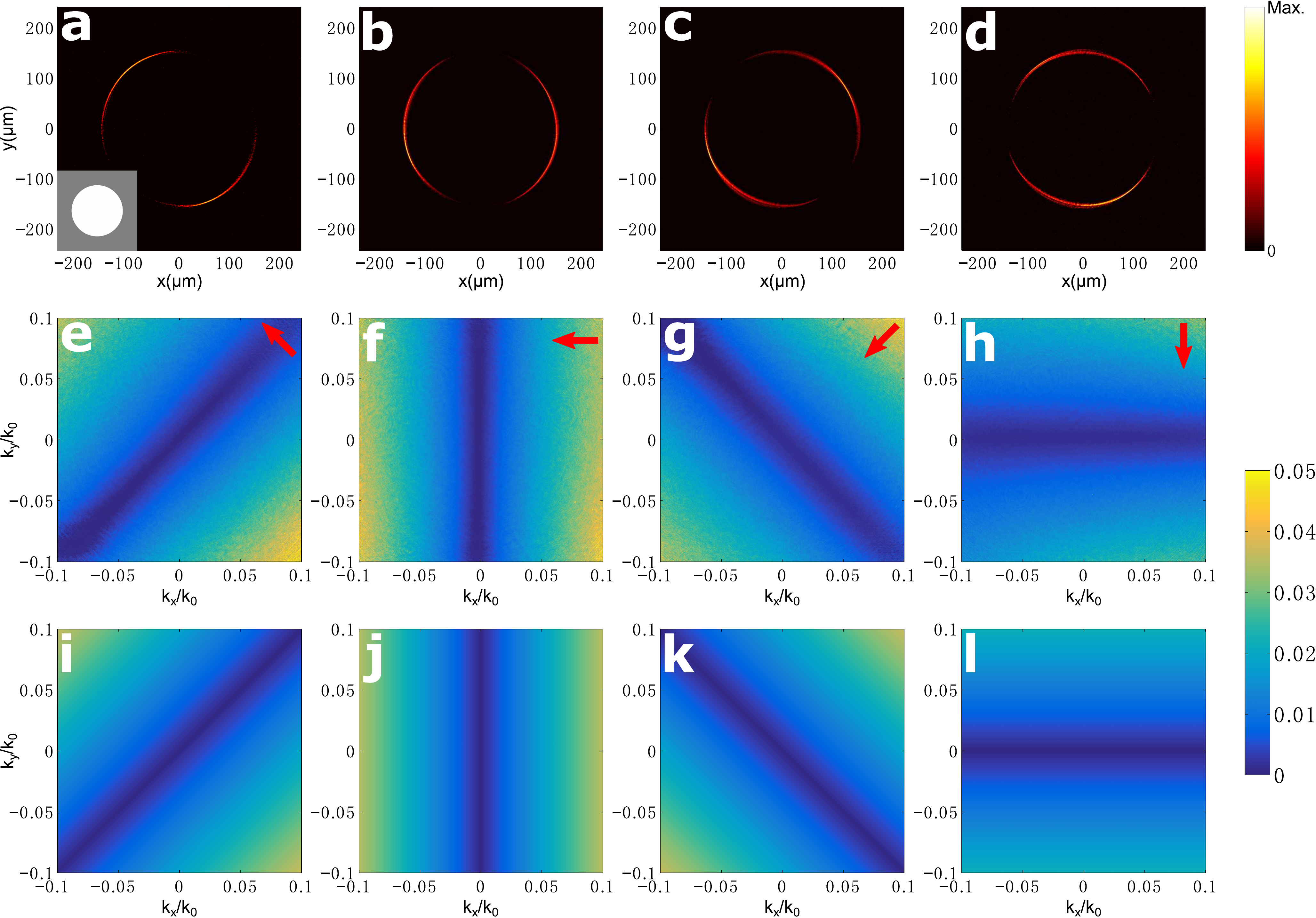}}%Here is how to import EPS art
\caption{\label{fig:3} Measurement of spatial differentiation results and corresponding spatial spectral transfer functions along different directions. (a-d) Measured spatial differentiation results of the phase distribution [the insert in (a)] along different directions with  $\varphi  = 45^\circ $,  $90^\circ $, $135^\circ $  and  $180^\circ $, corresponding to points c, d, e and f in Fig.~\ref{fig:2}(a), respectively. The insert in (a) is a disc test pattern with different phases for the gray and the white areas. (e-h) Experimental results of spatial spectral transfer functions, corresponding to (a-d). (i-l) Corresponding theoretical results calculated based on Eq.~(\ref{eq:3}).}
\end{figure}

As the experimental demonstration, Figs.~\ref{fig:3}(a-d) show the edge-enhanced differential contrast images for a disc phase distribution [shown as the insert in Fig.~\ref{fig:3}(a)], with  ${\gamma _1}$ and  ${\gamma _2}$ for the angles  of differentiation direction $\varphi  = 45^\circ $,  $90^\circ $, $135^\circ $  and  $180^\circ $. Figures~\ref{fig:3}(a-d) clearly exhibit the edges of the disc pattern as a circle, except the parts that are parallel to the differentiation direction. These results indeed show the adjustability of the direction of spatial differentiation.

For an accurate evaluation of the spatial differentiation direction, we experimentally measure the spatial spectral transfer functions as depicted in Figs.~\ref{fig:3}(e-h) (See the detailed procedures in Materials and Methods). As Eq.~(\ref{eq:3}) expects, Figs.~\ref{fig:3}(e-h) indeed exhibit linear dependences on both ${k_x}$  and ${k_y}$. For each case, the measured direction angles $\varphi $  are  $45.11^\circ $,  $90.01^\circ $, $134.91^\circ $  and  $180.00^\circ $ respectively, which are determined by fitting the gradient directions of the experimentally measured spatial spectral transfer functions (see Supplementary Section 2). The measured results  are shown as the arrows in Figs.~\ref{fig:3}(e-h), respectively. For comparison, the theoretical spatial spectral transfer functions are calculated from Eq.~(\ref{eq:3}) and shown as Figs.~\ref{fig:3}(i-l) respectively, which agree well with the experimental ones. We note that specifically when  ${\gamma _1} = 0^\circ $ or  $90^\circ $, the spatial differentiation is only from the IF shift and exhibits vertically as $\varphi  = 0^\circ $ or $180^\circ $.  It becomes totally horizontal for  ${\gamma _1} = {\gamma _2}$, where the spatial differentiation is purely induced by the angular GH shift since coefficient of the vertical spatial differentiation is  ${C_2} = 0$.

\begin{figure}[t]
\centerline{\includegraphics[width=3.2in]{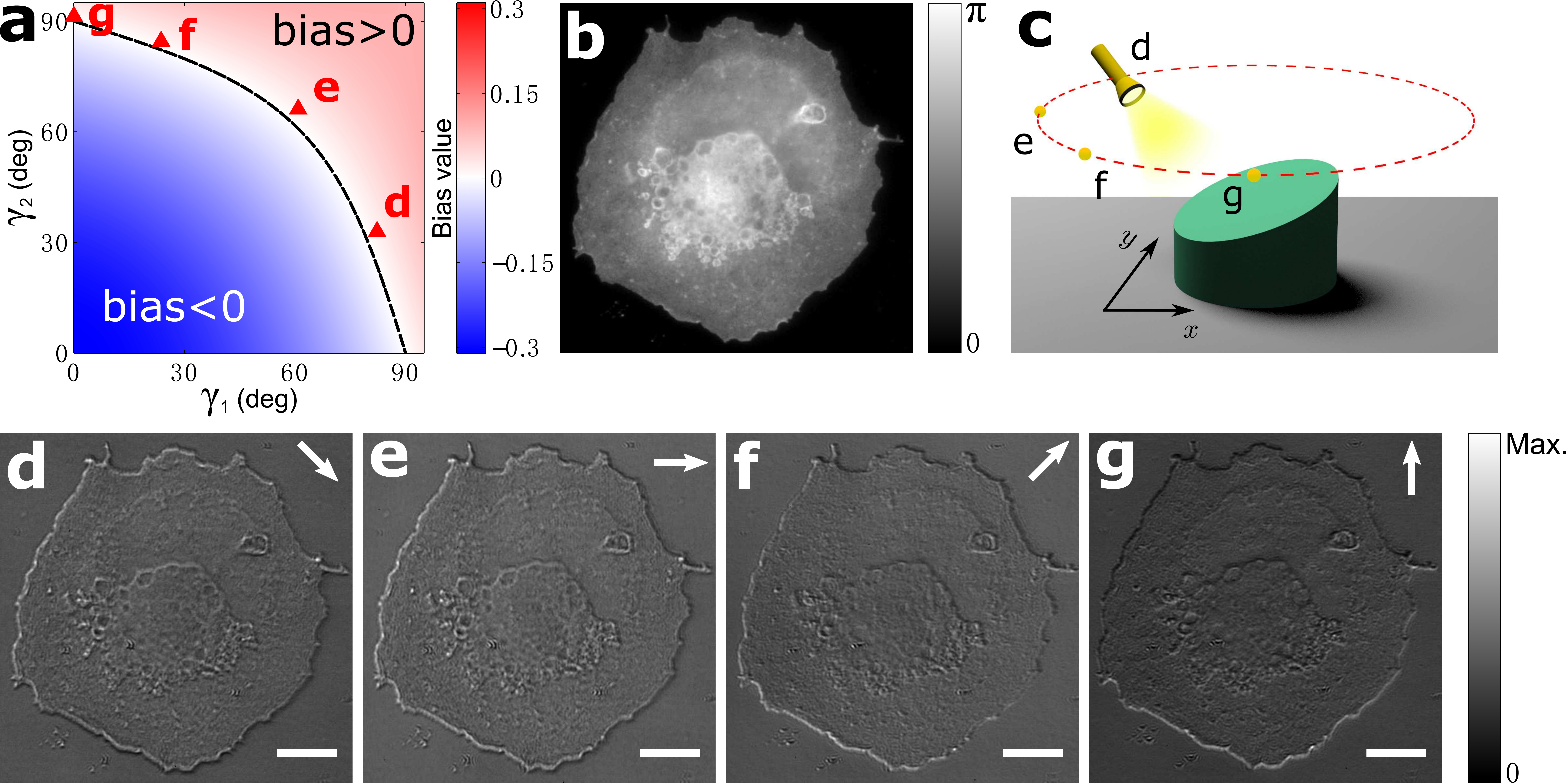}}%Here is how to import EPS art
\caption{\label{fig:4} Experimental demonstration of bias introduction and shadow-cast effect in differential contrast imaging of phase objects. (a) Theoretically calculated bias values under different ${\gamma _1}$  and  ${\gamma _2}$. The black dashed curve corresponds to the cross-polarization condition, where the bias value $b = 0$. (b) Phase distribution on the SLM, converted from an epithelial cell's image. (c) Schematic of a virtual light source obliquely illuminating an object (green). The shadow-cast effect with virtual illumination from the points d-g schematically correspond to those in the biased images (d-g). (d-g) Measured biased differential contrast images with positive bias values corresponding to the triangle points d-g in (a), with bias values  $b = 0.0163$,  $0.0151$, $0.0100$  and  $0.0054$, respectively. The white arrows indicate the orientations of the shadow cast in measured images. The white bars correspond to the length of $100\mu m$.}
\end{figure}

Next we experimentally demonstrate the biased imaging for a transparent phase object, by introducing a uniform constant background as the bias. According to Eq.~(\ref{eq:5}), we first simulate the bias value $b$ as shown in Fig.~\ref{fig:4}(a), where the black dashed line corresponds to  $b = 0$, i.e., the cross-polarization condition is satisfied. In order to demonstrate the biased effect, we generate the phase distribution of the incident light with a reflective phase-only spatial light modulator (SLM). An epithelial cell's image \cite{zeisscell} in Fig.~\ref{fig:4}(b) is loaded into SLM, with a prior quantitative phase distribution. We introduce the bias by only controlling the value of  ${\gamma _2}$ deviating from the black dashed line to red triangle points in Fig.~\ref{fig:4}(a). The specific deviation angles of  ${\gamma _2}$ are  $9^\circ $,  $5^\circ $,  $2^\circ $ and  $1^\circ $, corresponding to bias values  $b = 0.0163$,  $0.0151$, $0.0100$  and $0.0054$  respectively (see Supplementary Section 4). As the results, Figs.~\ref{fig:4}(d-g) show the biased contrast images, which seem like a virtual light source obliquely illuminates the object along different directions [schematically shown in Fig.~\ref{fig:4}(c)]. These shadow-cast images result from the positive biases which render the ridges and the troughs in phase distribution as the bright edges and the shadowed ones, respectively.

With a positive and very small bias as a perturbation, we can determine the sign of the spatial differentiation ${{\partial \psi } \mathord{\left/
 {\vphantom {{\partial \psi } {\partial l}}} \right.
 \kern-\nulldelimiterspace} {\partial l}}$. The absolute value of  ${{\partial \psi } \mathord{\left/
 {\vphantom {{\partial \psi } {\partial l}}} \right.
 \kern-\nulldelimiterspace} {\partial l}}$ can be obtained from the non-bias case, and therefore we experimentally acquire the first-order directional derivatives of  $\psi \left( {x,y} \right)$ (see the details in Supplementary Section 5). Figures~\ref{fig:5}(a) and ~\ref{fig:5}(b) show two directional derivatives of the incident phase distribution shown as Fig.~\ref{fig:4}(b) along $y$- and $x$-direction, respectively. For comparison, the ideal spatial differentiation along $y$- and $x$-direction are calculated as Figs.~\ref{fig:5}(c) and ~\ref{fig:5}(d), respectively. These experimental results clearly coincide well with the calculated ones, indicating great accuracy of the performance.  Finally we recover the phase from the obtained directional derivatives through the 2D Fourier algorithm. Figure~\ref{fig:5}(e) shows the result of the recovered phase distribution, which coincides with the original incident one [Fig.~\ref{fig:5}(f)]. The normalized mean square error (NMSE) between the recovered result and the original distribution is calculated as 8.82$\%$, which can be further reduced with an optimized image system. In this way, we not only enhance the contrast to make transparent objects visible, but successfully recover its original phase distribution.

 \begin{figure}[t]
\centerline{\includegraphics[width=3.2in]{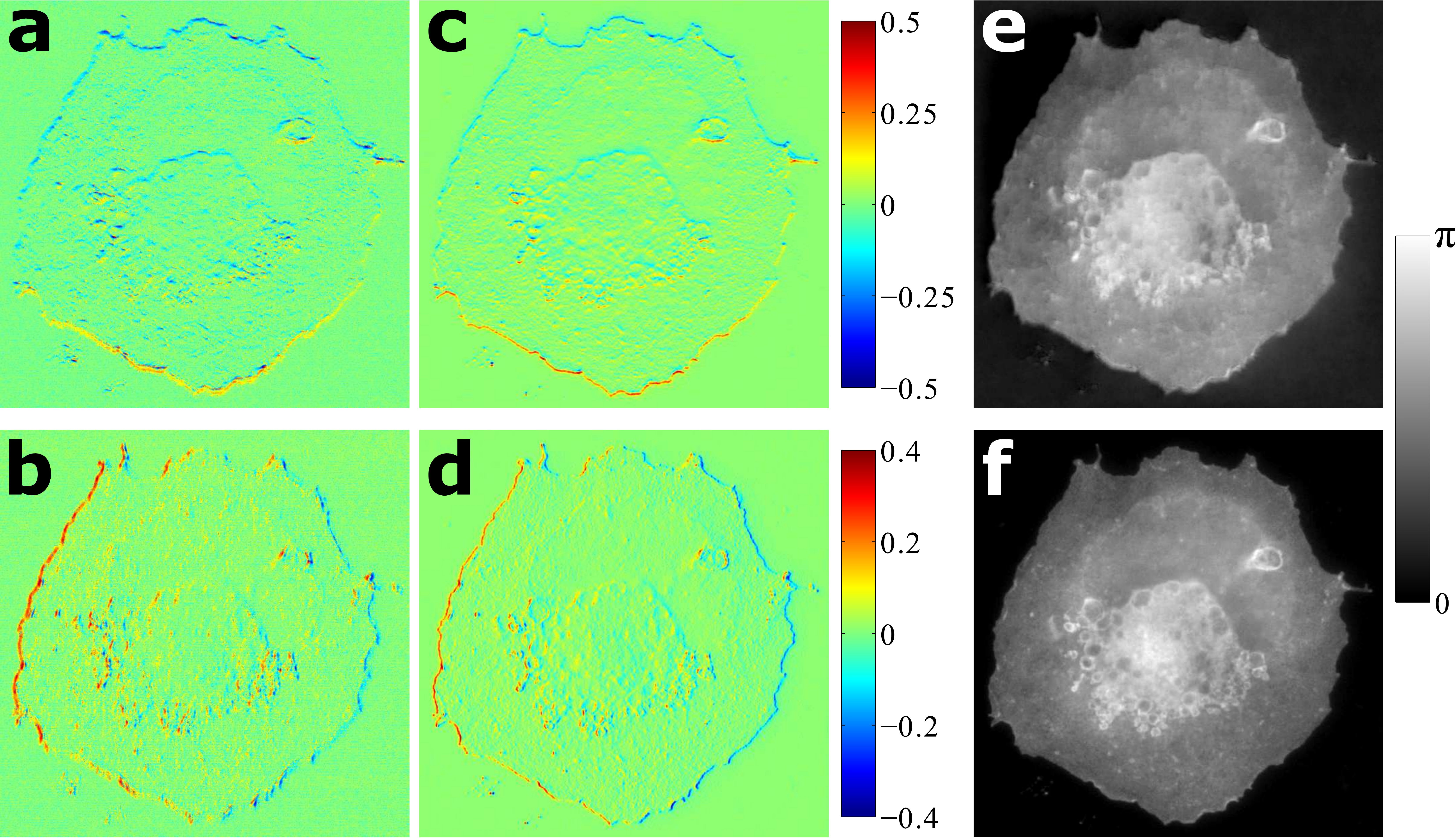}}%Here is how to import EPS art
\caption{\label{fig:5} Directional derivatives of phase distribution in Fig.~\ref{fig:4}(b) and the corresponding recovered phase. (a, b) Experimental results of vertical and horizontal partial derivatives of phase distribution in Fig.~\ref{fig:4}(b), respectively. (c, d) Ideal vertical and horizontal partial derivatives, respectively. (e) Recovered phase distribution from (a) and (b) using the 2D Fourier algorithm. (f) Original phase distribution on the SLM [the same as Fig.~\ref{fig:4}(b)].}
\end{figure}

\section{Discussion}

We experimentally demonstrate a phase mining method only by analyzing the polarization in light reflection on a single planar dielectric interface. The direction of the spatial differentiation is continuously adjustable assuring differential contrast enhancement along different directions. Importantly, the method easily introduces a bias and generates shadow-cast differential contrast images. Based on this bias scheme, we determine the sign of the spatial differentiation of the phase distribution and finally recover the original phase information for a uniform-intensity image.

 We note that the present method is also suitable for birefringent specimens with advantages over conventional DIC, because the illumination happens before polarization analysis in light reflection on a dielectric interface. This planar-interface scheme is much simpler and almost costless in comparison to the methods with complex Nomarski prisms or spatial light modulators. This method also works under a partially coherent illumination. For example, here a partially coherent beam is used for illumination of the phase object, in order to eliminate speckles and enhance the image quality. Our experimental results confirm that the proposed method is material-in-general and therefore by X-ray or electron polarization analysis \cite{belyakov1989polarization, scheinfein1990scanning} it opens a new avenue to quantify the phase through X-ray or electron microscopy image.

\section{Materials and methods}
\subsection{Experimental measurement of spatial spectral transfer functions}

We experimentally measure the spatial spectral transfer functions of different directional spatial differentiations under the cross-polarization condition. Specifically, we measure the incident and reflected spatial spectra, and the spatial spectral transfer functions are obtained by normalizing the reflected spectrum data with the incident ones. The experiment setups for measuring the spatial spectral transfer functions is schematically shown in Fig.~ S1 of Supplementary Material. A green laser is collimated to a Gaussian beam. Then we use a lens L1 to focus the incident beam to amplify the amplitudes of the plane wave components with higher spatial frequencies, which is advantageous for the measurement of spatial spectral transfer function with a relatively broad spatial bandwidth. The beam is subsequently polarized by a polarizer P1 with an orientation angle ${\gamma _1}$  as shown in Fig.~ S1(a) and focuses on a dielectric interface. e.g. an air-glass interface in our experiment. After reflected on the air-glass interface, the beam is analyzed by another polarizer P2 whose orientation angle is ${\gamma _2}$,  determined by the cross-polarization condition. Another lens L2 performs Fourier transform of the reflected field and the corresponding reflected spatial spectra is measured by a beam profiler (Ophir SP620U) at the back focal plane of L2. On the other hand, the incident spatial spectra are obtained by removing the glass and rotating the reflection path (lens L2 and polarizer P2) as shown in Fig.~ S1(b), while the orientation angles of P2 are rotated to be the same as those of P1. We note that the two polarizers are placed between lenses L1 and L2 in order to avoid the polarization rotation induced by the geometric phase during light focusing and collimating \cite{Zhu2019Generalized}. By normalizing the reflected spectrum data with the incident ones, we can obtain the spatial spectral transfer functions of the adjustable spatial differentiation.
%\bibliographystyle{osajnl}
%\bibliography{./References}

\clearpage
\newpage

\appendix

%\makeatletter
%\renewcommand{\thefigure}{S\arabic{figure}}
%\def\tagform@#1{\maketag@@@{(\ignorespaces\textbf{S#1}\unskip\@@italiccorr)}}
%\renewcommand{\eqref}[1]{\textup{{\normalfont S\ref{#1}}\normalfont}}
%\makeatother

\newcommand{\hbAppendixPrefix}{S}
\renewcommand{\thefigure}{\hbAppendixPrefix\arabic{figure}}
\setcounter{figure}{0}

\renewcommand{\thetable}{\hbAppendixPrefix\arabic{table}}
\setcounter{table}{0}
\renewcommand{\theequation}{\hbAppendixPrefix\arabic{equation}}
\setcounter{equation}{0}

\onecolumngrid
%=============================================================================
\begin{center}
\large{\textbf{
Supplementary Information}}
\end{center}

\subsection*{Calculation of the spatial spectral transfer function during a reflection process}	

To calculate the spatial spectral transfer function, we first consider a plane wave with a wavevector $\vec k$ between the incident electric field ${\vec E_i}$ and the reflected one ${\vec E_r}$ on the dielectric interface. We define the s- and p-polarizations as the electric field vectors along ${\vec u_s} = {{{{\vec u}_k} \times \vec n} \mathord{\left/
 {\vphantom {{{{\vec u}_k} \times \vec n} {\left| {{{\vec u}_k} \times \vec n} \right|}}} \right.
 \kern-\nulldelimiterspace} {\left| {{{\vec u}_k} \times \vec n} \right|}}$ and ${\vec u_p} = {\vec u_s} \times {\vec u_k}$, respectively. Here ${\vec u_k} = {{\vec k} \mathord{\left/
 {\vphantom {{\vec k} {\left| {\vec k} \right|}}} \right.
 \kern-\nulldelimiterspace} {\left| {\vec k} \right|}}$ is the normalized wavevector and $\vec n$ is the unit normal vector of the interface. Then, the left- and right-handed circular polarization bases ${\vec u_ + }$ and ${\vec u_ - }$ for plane waves are defined as
\begin{equation}
\begin{aligned}
{\vec u_ \pm } = \frac{1}{{\sqrt 2 }}\left( {{{\vec u}_p} \pm i{{\vec u}_s}} \right).
\label{eq:S1}
\end{aligned}
\end{equation}
We can decompose the vectorial fields ${\vec E_i}$ and ${\vec E_r}$ into left- and right-handed circularly polarized plane waves through the spatial Fourier transform:
\begin{subequations}
\begin{equation}
\begin{aligned}
{{\vec E}_i} &= \vec u_ + ^{i0}E_ + ^{i0}\left( {x,y} \right) + \vec u_ - ^{i0}E_ - ^{i0}\left( {x,y} \right) + \vec u_z^{i0}E_z^{i0}\left( {x,y} \right)\\
 &= \iint {\left[ {\vec u_ + ^{i0}\tilde E_ + ^{i0}\left( {{k_x},{k_y}} \right) + \vec u_ - ^{i0}\tilde E_ - ^{i0}\left( {{k_x},{k_y}} \right) + \vec u_z^{i0}\tilde E_z^{i0}\left( {{k_x},{k_y}} \right)} \right]{e^{i{k_x}}}{e^{i{k_y}}}d{k_x}d{k_y}},
\label{eq:S2a}
\end{aligned}
\end{equation}
\begin{equation}
\begin{aligned}
{{\vec E}_r} &= \vec u_ + ^{r0}E_ + ^{r0}\left( {x,y} \right) + \vec u_ - ^{r0}E_ - ^{r0}\left( {x,y} \right) + \vec u_z^{r0}E_z^{r0}\left( {x,y} \right)\\
 &= \iint {\left[ {\vec u_ + ^{r0}\tilde E_ + ^{r0}\left( {{k_x},{k_y}} \right) + \vec u_ - ^{r0}\tilde E_ - ^{r0}\left( {{k_x},{k_y}} \right) + \vec u_z^{r0}E_z^{r0}\left( {{k_x},{k_y}} \right)} \right]{e^{i{k_x}}}{e^{i{k_y}}}d{k_x}d{k_y}}.
\label{eq:S2b}
\end{aligned}
\end{equation}
\end{subequations}
Here, $x$ and $y$ are the beam coordinates as shown in Fig.~1 in the main text, while $z$ is along the propagation direction of the beam and $\vec u_z^{i\left( r \right)}$ is the corresponding unit vector. $\vec u_ \pm ^{i0}$ and $\vec u_ \pm ^{r0}$ are the circular polarization bases for the central wavevector of the incident and reflected beams, respectively. For calculation of the spatial spectra transfer function, we transfer the electric field of each plane wave to the polarization basis for its own wavevector:
\begin{subequations}
\begin{equation}
\begin{aligned}
\left( {\begin{array}{*{20}{c}}
{\tilde E_ + ^i}\\
{\tilde E_ - ^i}
\end{array}} \right) = {U_1}\left( {\begin{array}{*{20}{c}}
{\tilde E_ + ^{i0}}\\
\begin{array}{l}
\tilde E_ - ^{i0}\\
\tilde E_z^{i0}
\end{array}
\end{array}} \right),
\label{eq:S3a}
\end{aligned}
\end{equation}
\begin{equation}
\begin{aligned}
\left( {\begin{array}{*{20}{c}}
{\tilde E_ + ^r}\\
{\tilde E_ - ^r}
\end{array}} \right) = {U_2}\left( {\begin{array}{*{20}{c}}
{\tilde E_ + ^{r0}}\\
\begin{array}{l}
\tilde E_ - ^{r0}\\
\tilde E_z^{r0}
\end{array}
\end{array}} \right).
\label{eq:S3b}
\end{aligned}
\end{equation}
\end{subequations}
The matrices ${U_1}$ and ${U_2}$ are originated from the rotations of coordinates, describing the transformation between the circular polarization basis for the central wavevector and that for each other wavevector:
\begin{subequations}
\begin{equation}
\begin{aligned}
{U_1} = \left( {\begin{array}{*{20}{c}}
{{{\left( {\vec u_ + ^i} \right)}^ * } \cdot \vec u_ + ^{i0}}&{{{\left( {\vec u_ + ^i} \right)}^ * } \cdot \vec u_ - ^{i0}}&{{{\left( {\vec u_ + ^i} \right)}^ * } \cdot \vec u_z^{i0}}\\
{{{\left( {\vec u_ - ^i} \right)}^ * } \cdot \vec u_ + ^{i0}}&{{{\left( {\vec u_ - ^i} \right)}^ * } \cdot \vec u_ - ^{i0}}&{{{\left( {\vec u_ - ^i} \right)}^ * } \cdot \vec u_z^{i0}}
\end{array}} \right),
\label{eq:S4a}
\end{aligned}
\end{equation}
\begin{equation}
\begin{aligned}
{U_2} = \left( {\begin{array}{*{20}{c}}
{{{\left( {\vec u_ + ^r} \right)}^ * } \cdot \vec u_ + ^{r0}}&{{{\left( {\vec u_ + ^r} \right)}^ * } \cdot \vec u_ - ^{r0}}&{{{\left( {\vec u_ + ^r} \right)}^ * } \cdot \vec u_z^{r0}}\\
{{{\left( {\vec u_ - ^r} \right)}^ * } \cdot \vec u_ + ^{r0}}&{{{\left( {\vec u_ - ^r} \right)}^ * } \cdot \vec u_ - ^{r0}}&{{{\left( {\vec u_ - ^r} \right)}^ * } \cdot \vec u_z^{r0}}
\end{array}} \right).
\label{eq:S4b}
\end{aligned}
\end{equation}
\end{subequations}
Due to the continuous condition of the tangential wavevector along the interface, the incident plane wave with the transverse component $\left( {{k_x},{k_y}} \right)$ only generates the reflected plane wave with the same $\left( {{k_x},{k_y}} \right)$. Therefore, for each incident plane wave with wavevector  $\vec k$, the Fourier spectra of the reflected plane wave is
\begin{equation}
\begin{aligned}
\left( {\begin{array}{*{20}{c}}
{\tilde E_ + ^{r0}}\\
\begin{array}{l}
\tilde E_ - ^{r0}\\
\tilde E_z^{r0}
\end{array}
\end{array}} \right) = R\left( {\begin{array}{*{20}{c}}
{\tilde E_ + ^{i0}}\\
\begin{array}{l}
\tilde E_ - ^{i0}\\
\tilde E_z^{i0}
\end{array}
\end{array}} \right) = U_2^\dag \tilde R{U_1}\left( {\begin{array}{*{20}{c}}
{\tilde E_ + ^{i0}}\\
\begin{array}{l}
\tilde E_ + ^{i0}\\
\tilde E_z^{i0}
\end{array}
\end{array}} \right).
\label{eq:S5}
\end{aligned}
\end{equation}
Here, matrix $R$ describes the reflection coefficients for left- and right-handed circularly polarized plane waves:
\begin{equation}
\begin{aligned}
\tilde R = \frac{1}{2}\left( {\begin{array}{*{20}{c}}
{{r_p} + {r_s}}&{{r_p} - {r_s}}\\
{{r_p} - {r_s}}&{{r_p} + {r_s}}
\end{array}} \right),
\label{eq:S6}
\end{aligned}
\end{equation}
where ${r_p}$ and ${r_s}$ are the Fresnel's reflection coefficients for each p- and s-polarized plane waves with the wavevector $\vec k$, respectively.

Under the paraxial approximation where both the incident and reflected field dominate in the transversal component and $\tilde E_z^{i0\left( {r0} \right)} \approx 0$, the dimensions of matrices $R$, ${U_1}$ and ${U_2}$ can be reduced to $2 \times 2$. Specifically, the transfer matrices  ${U_1}$ and ${U_2}$  can be approximately evaluated through the geometric phases for left- and right-handed circularly polarized waves as
\begin{equation}
\begin{aligned}
{U_j} = \left( {\begin{array}{*{20}{c}}
{\exp \left( {i\Phi _B^j} \right)}&0\\
0&{\exp \left( { - i\Phi _B^j} \right)}
\end{array}} \right),
\label{eq:S7}
\end{aligned}
\end{equation}
where $j = 1,2$. $\Phi _B^1 = \frac{{{k_y}}}{{{k_0}}}\cot {\theta _0}$ and $\Phi _B^2 =  - \frac{{{k_y}}}{{{k_0}}}\cot {\theta _0}$ are the geometric phases during the wavevector rotation \cite{bliokh2013goos}, while ${\theta _0}$ is the incident angle of the plane wave component with the central wavevector. Equation~(\ref{eq:S7}) shows that the left- and right-handed circularly polarized light with a wavevector  $\vec k$ experience the opposite geometric phases during the light reflection.

As shown in Fig.~1 of the main text, the vectorial electric field after the first polarizer P1 can be written as ${\vec E_i} = {S_{in}}\left( {x,y} \right){\vec V_1}$, where  ${S_{in}}\left( {x,y} \right)$ is the incident field distribution and  ${\vec V_1} = \frac{i}{{\sqrt 2 }}{\left( { - {e^{i{\gamma _1}}},{e^{ - i{\gamma _1}}},0} \right)^T}$ is the unit vector determined by the orientation of P1 under the circular polarization basis of the central wavevector. Then after the reflection on an air-glass interface and analyzed by the second polarizer P2, the measured field distribution is ${S_{out}}\left( {x,y} \right) = \vec V_2^\dag  \cdot {\vec E_r}$ where  ${\vec V_2} = \frac{i}{{\sqrt 2 }}{\left( { - {e^{i{\gamma _2}}},{e^{ - i{\gamma _2}}},0} \right)^T}$. By a spatial Fourier transform, ${S_{in\left( {out} \right)}}$ can also be written as the superposition of plane waves as ${S_{in\left( {out} \right)}} = \iint {{{\tilde S}_{in\left( {out} \right)}}\left( {{k_x},{k_y}} \right)\exp \left( {i{k_x}x} \right)\exp \left( {i{k_y}y} \right)d{k_x}d{k_y}}$. Thus, we have
\begin{subequations}
\begin{equation}
\begin{aligned}
\left( {\begin{array}{*{20}{c}}
{\tilde E_ + ^{i0}}\\
\begin{array}{l}
\tilde E_ - ^{i0}\\
\tilde E_z^{i0}
\end{array}
\end{array}} \right) = {\tilde S_{in}}\left( {x,y} \right){\vec V_1},
\label{eq:S8a}
\end{aligned}
\end{equation}
\begin{equation}
\begin{aligned}
{\tilde S_{out}}\left( {x,y} \right) = \vec V_2^\dag  \cdot \left( {\begin{array}{*{20}{c}}
{\tilde E_ + ^{r0}}\\
\begin{array}{l}
\tilde E_ - ^{r0}\\
\tilde E_z^{r0}
\end{array}
\end{array}} \right).
\label{eq:S8b}
\end{aligned}
\end{equation}
\end{subequations}
By substituting Eq.~(\ref{eq:S8a}) and (\ref{eq:S8b}), Eq.~(\ref{eq:S5}) becomes
\begin{equation}
\begin{aligned}
{\vec V_2}{\tilde S_{out}}\left( {{k_x},{k_y}} \right) = R{\vec V_1}{\tilde S_{in}}\left( {{k_x},{k_y}} \right).
\label{eq:S9}
\end{aligned}
\end{equation}
According to Eq.~(\ref{eq:S9}), the spatial spectral transfer function between the specimen's distribution and the finally measured one is
\begin{equation}
\begin{aligned}
H \equiv \frac{{{{\tilde S}_{out}}\left( {{k_x},{k_y}} \right)}}{{{{\tilde S}_{in}}\left( {{k_x},{k_y}} \right)}} = \vec V_2^\dag R{\vec V_1}.
\label{eq:S10}
\end{aligned}
\end{equation}
By calculation, the spatial spectral transfer function $H$ in Eq.~(\ref{eq:S10}) is
\begin{equation}
\begin{aligned}
H = {r_p}\sin {\gamma _1}\sin {\gamma _2} + {r_s}\cos {\gamma _1}\cos {\gamma _2} + \frac{{\delta \left( {{r_p} + {r_s}} \right)\sin \left( {{\gamma _2} - {\gamma _1}} \right)}}{2}{k_y},
\label{eq:S11}
\end{aligned}
\end{equation}
where $\delta  = \frac{2}{{{k_0}}}\cot {\theta _0}$  corresponds to the transverse shift for the circularly polarized plane waves. We note that the two items in Eq.~(\ref{eq:S11}) exactly correspond to the Malus's Law in the configuration of Fig.~1, while the third item is due to the Imbert-Fedorov (IF) shift.

\subsection*{Experimental measurement of the differentiation direction angle $\varphi $  and the proportionality coefficient $A$ for different combination of ${\gamma _1}$  and ${\gamma _2}$}

\begin{figure}[h]
\setcaptionwidth{6in}
\centerline{\includegraphics[width=4.5in]{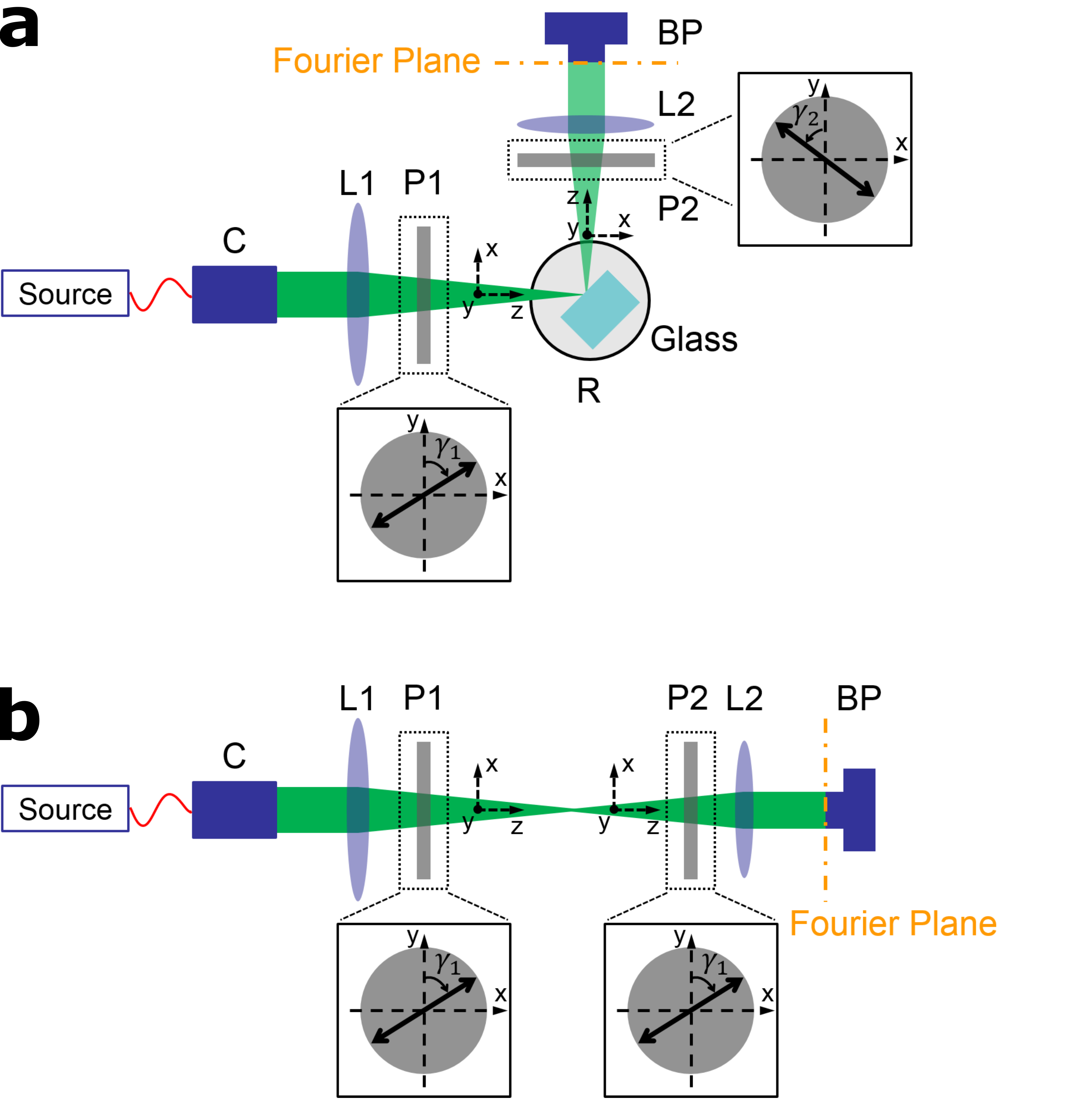}}%Here is how to import EPS art
\caption{\label{fig:S2} Experimental setup for measuring the spatial spectral transfer function. (a) Experimental setup for measuring the reflected spatial spectrum: collimator C; lenses L1 and L2 with focal lengths $50mm$ and $30mm$, respectively; polarizers P1 and P2; glass slab (material BK7); precision rotator R; and beam profiler BP (Ophir SP620U). A green laser (wavelength  ${\lambda _0} = 532{\rm{nm}}$) is connected to the collimator through a fiber with a polarization controller. (b) The experimental setup for measuring the incident spatial spectrum, in which the polarizer P2 is oriented with the same angle of P1.}
\end{figure}

To verify the theoretical result of Eq.~(3) in the main text, here we show the experimental measurement of the differentiation direction angle $\varphi $ and the proportionality coefficient $A$ for different combinations of ${\gamma _1}$  and  ${\gamma _2}$. Experimentally, we acquire the direction angles and proportionality coefficients by measuring the spatial spectral transfer functions [Fig.~\ref{fig:S2}]. Their gradient directions show the directions of spatial differentiation while gradient values indicate the proportionality coefficients. Specifically, we obtain the values of  ${C_1}$ and  ${C_2}$, by fitting the horizontal and vertical gradients of the measured spatial spectral transfer functions, respectively. Then we calculate the direction angles and proportionality coefficients based on the obtained values of ${C_1}$  and  ${C_2}$.

\begin{figure}[b]
\setcaptionwidth{6in}
\centerline{\includegraphics[width=6in]{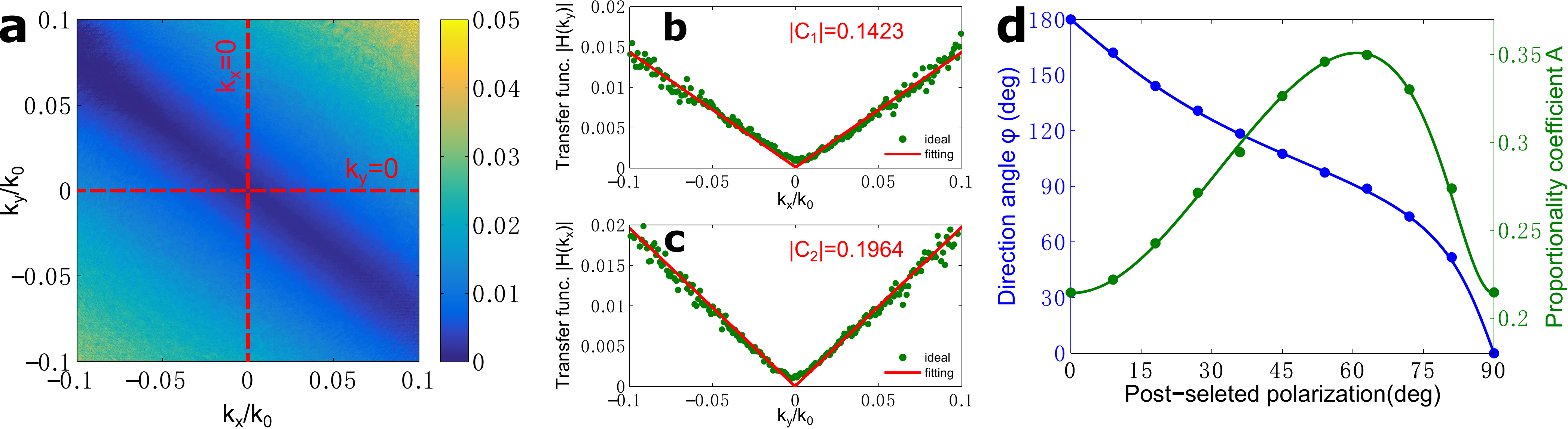}}%Here is how to import EPS art
\caption{\label{fig:S1} Experimental measurement of direction angle  $\varphi $ and the proportional coefficient $A$. In these results, the incident angle of the beam is  ${\theta _0} = 45^\circ $ and a BK7 glass slab works as the reflector with refractive index 1.5195 at the wavelength ${\lambda _0} = 532{\rm{nm}}$. (a) Experimentally measured spatial spectral transfer function with  ${\gamma _1} = 18.0^\circ $ and ${\gamma _2} = 84.23^\circ $. (b, c) Fitting results of the data on ${k_x} = 0$  and ${k_y} = 0$ [red dashed lines in (a)], respectively. (d) Values of  $\varphi $ (blue) and  $A$ (green) under different combinations of  ${\gamma _1}$ and  ${\gamma _2}$. The solid and dotted lines are the theoretical and experimental results, respectively.}
\end{figure}

As shown in Fig.~\ref{fig:S1}, it corresponds to a measured spatial spectral transfer function with  ${\gamma _1} = 18.0^\circ $ and  ${\gamma _2} = 84.23^\circ $, which satisfies the cross-polarization condition. We extract the data on ${k_x} = 0$  and  ${k_y} = 0$ [the red dashed lines in Fig.~\ref{fig:S1}(a)] and obtain the values of  ${C_1}$ and ${C_2}$  by fitting their slopes, respectively. In Figs.~\ref{fig:S1}(b) and ~\ref{fig:S1}(c), the data exhibit linear dependences on  ${k_x}$ and  ${k_y}$, respectively, and the fitting results are $\left| {{C_1}} \right| = 0.1423$  and  $\left| {{C_2}} \right| = 0.1964$. With $\varphi  \in \left[ {0,180} \right]$  and Fig.~\ref{fig:S1}(a) showing a rough gradient direction, we have ${C_1} =  - 0.1423$  and  ${C_2} =  - 0.1964$, respectively. According to the theoretical spatial spectral transfer function in Eq.~(3) in the main text, the direction and proportionality coefficient of the spatial differentiation are  $\vec l = \left( {{C_1},{C_2}} \right)$ and  $A = \left| {\vec l} \right| = \sqrt {{{\left( {{C_1}} \right)}^2} + {{\left( {{C_2}} \right)}^2}} $, respectively. Therefore, in the case of Fig.~\ref{fig:S1}(a), we have $\varphi  = 144.08^\circ $  and  $A = 0.2425$.

Following the above method, we experimentally measure the results under different combinations of ${\gamma _1}$  and  ${\gamma _2}$, among which four specific cases ($\varphi  = 45^\circ $,  $90^\circ $,  $135^\circ $ and  $180^\circ $) are shown in the main text. The experimental results are plotted as the dotted lines in Fig.~\ref{fig:S1}(d) and coincide well with the theoretical results, which are calculated from Eq.~(3) and shown as the solid lines. The results again exhibit the adjustability of the proposed spatial differentiation. These coefficients are quantitatively important. For instance, differentiated results along different directions should be first normalized by their corresponding proportionality coefficients before being utilized with each other in the meantime.

\subsection*{Field transformation in the bias introduction scheme}

In the biased condition  ${r_{p0}}\sin {\gamma _1}\sin {\gamma _2} =  - {r_{s0}}\cos {\gamma _1}\cos {\gamma _2} + b$, the spatial spectral transfer function Eq.~(\ref{eq:S11}) becomes
\begin{equation}
\begin{aligned}
H =  - \left( {{C_1}{k_x} + {C_2}{k_y}} \right) + b.
\label{eq:S12}
\end{aligned}
\end{equation}
In the spatial domain, the measured field distribution is
\begin{equation}
\begin{aligned}
{S_{out}}\left( {x,y} \right) = i\left( {{C_1}\frac{{\partial {S_{in}}}}{{\partial x}} + {C_2}\frac{{\partial {S_{in}}}}{{\partial y}}} \right) + b{S_{in}}
 = iA\frac{{\partial {S_{in}}}}{{\partial l}} + b{S_{in}}.
\label{eq:S13}
\end{aligned}
\end{equation}
For an observed field with a phase distribution  $\psi \left( {x,y} \right)$, the incident field is  ${S_{in}} = {e^{ - i\psi \left( {x,y} \right)}}$. According to Eq.~(\ref{eq:S13}), we have the output field written as
\begin{equation}
\begin{aligned}
{S_{out}}\left( {x,y} \right) = A\frac{{\partial \psi }}{{\partial l}}{e^{ - i\psi \left( {x,y} \right)}} + b{e^{ - i\psi \left( {x,y} \right)}} = \left( {A\frac{{\partial \psi }}{{\partial l}} + b} \right){e^{ - i\psi \left( {x,y} \right)}}.
\label{eq:S14}
\end{aligned}
\end{equation}
By direct intensity detection, the result is measured as the absolute value  $\left| {A\frac{{\partial \psi }}{{\partial l}} + b} \right|$. As a consequence, with an appropriate positive bias  $b$, the measured distribution appears brighter or darker at areas with positive or negative directional derivative  ${{\partial \psi } \mathord{\left/
 {\vphantom {{\partial \psi } {\partial l}}} \right.
 \kern-\nulldelimiterspace} {\partial l}}$, respectively.

\subsection*{Specific orientation angles of polarizers P1 and P2 for different directional spatial differentiations and biased images}

\begin{table}[h]
	\caption{}\label{table:S1}	
	\centering
	\begin{subtable}[t]{5in}
		\centering
		\caption{Orientation angles ${\gamma _1}$  and  ${\gamma _2}$ of polarizers for spatial differentiation with different direction angle  $\varphi $, and the corresponding bias values}\label{table:1a}
		\begin{tabular}{|p{1.5cm}<{\centering}|p{2.5cm}<{\centering}|p{2.5cm}<{\centering}|p{2.5cm}<{\centering}|p{2.5cm}<{\centering}|}
		\hline
		 & Figure~3(a) & Figure~3(b) & Figure~3(c) & Figure~3(d) \\
		\hline
		$\varphi $ & $45^\circ $ & $90^\circ $ & $135^\circ $ & $180^\circ $ \\
        \hline
        ${\gamma _1}$ & $82.25^\circ $ & $60.86^\circ $ & $23.65^\circ $ & $0^\circ $ \\
        \hline
        ${\gamma _2}$ & $23.65^\circ $ & $60.86^\circ $ & $82.25^\circ $ & $90^\circ $ \\
        \hline
        $b$ & $0$ & $0$ & $0$ & $0$ \\
		\hline
		\end{tabular}
	\end{subtable}
	
	\begin{subtable}[t]{5in}
		\centering
		\caption{Orientation angles  ${\gamma _1}$  and  ${\gamma _2}$  of polarizers for biased differential contrast images with different bias values  $b$}\label{table:1a}
		\begin{tabular}{|p{1.5cm}<{\centering}|p{2.5cm}<{\centering}|p{2.5cm}<{\centering}|p{2.5cm}<{\centering}|p{2.5cm}<{\centering}|}
		\hline
		 & Figure~4(a) & Figure~4(b) & Figure~4(c) & Figure~4(d) \\
		\hline
		$\varphi $ & $45^\circ $ & $90^\circ $ & $135^\circ $ & $180^\circ $ \\
        \hline
        ${\gamma _1}$ & $82.25^\circ $ & $60.86^\circ $ & $23.65^\circ $ & $0^\circ $ \\
        \hline
        ${\gamma _2}$ & $32.65^\circ $ & $65.86^\circ $ & $84.25^\circ $ & $91^\circ $ \\
        \hline
        $b$ & $0.0163$ & $0.0151$ & $0.0100$ & $0.0054$ \\
		\hline
		\end{tabular}
	\end{subtable}
\end{table}

The orientation angles  ${\gamma _1}$ and ${\gamma _2}$  for a certain direction angle $\varphi $  can be determined according to Eq.~(2). Figure~2(b) in the main text plots the values of  ${\gamma _1}$ and ${\gamma _2}$ under every  $\varphi $ for a specific case, in which the wavelength of laser source is $\lambda  = 532nm$  and the corresponding refractive index of the BK7 glass is 1.5195. For examples, the values of ${\gamma _1}$  and  ${\gamma _2}$ for the direction angles  $\varphi  = 45^\circ $,  $90^\circ $, $135^\circ $  and $180^\circ $  are listed in Table~\ref{table:S1}(a). In these cases, the cross-polarization condition is critically satisfied so that the corresponding bias values are $b = 0$.

The biased differential contrast images can be obtained just by controlling the value of ${\gamma _2}$  deviating from the cross-polarization condition. For example, we show biased images with different bias values in the main text [Figs.~4(d-g)]. We keep the values of orientation angle   the same as in Table~\ref{table:S1}(a) and rotate polarizer P2 for deviation angles  $9^\circ $,  $5^\circ $, $2^\circ $  and  $1^\circ $, respectively. The new values of ${\gamma _2}$  and the corresponding bias values $b$  are listed in the Table~\ref{table:S1}(b). These four cases correspond to biased differential contrast images Figs.~4(d-g) in the main text.

\subsection*{Acquisition of directional derivatives of the phase distribution}

\begin{figure}[h]
\setcaptionwidth{6in}
\centerline{\includegraphics[width=6in]{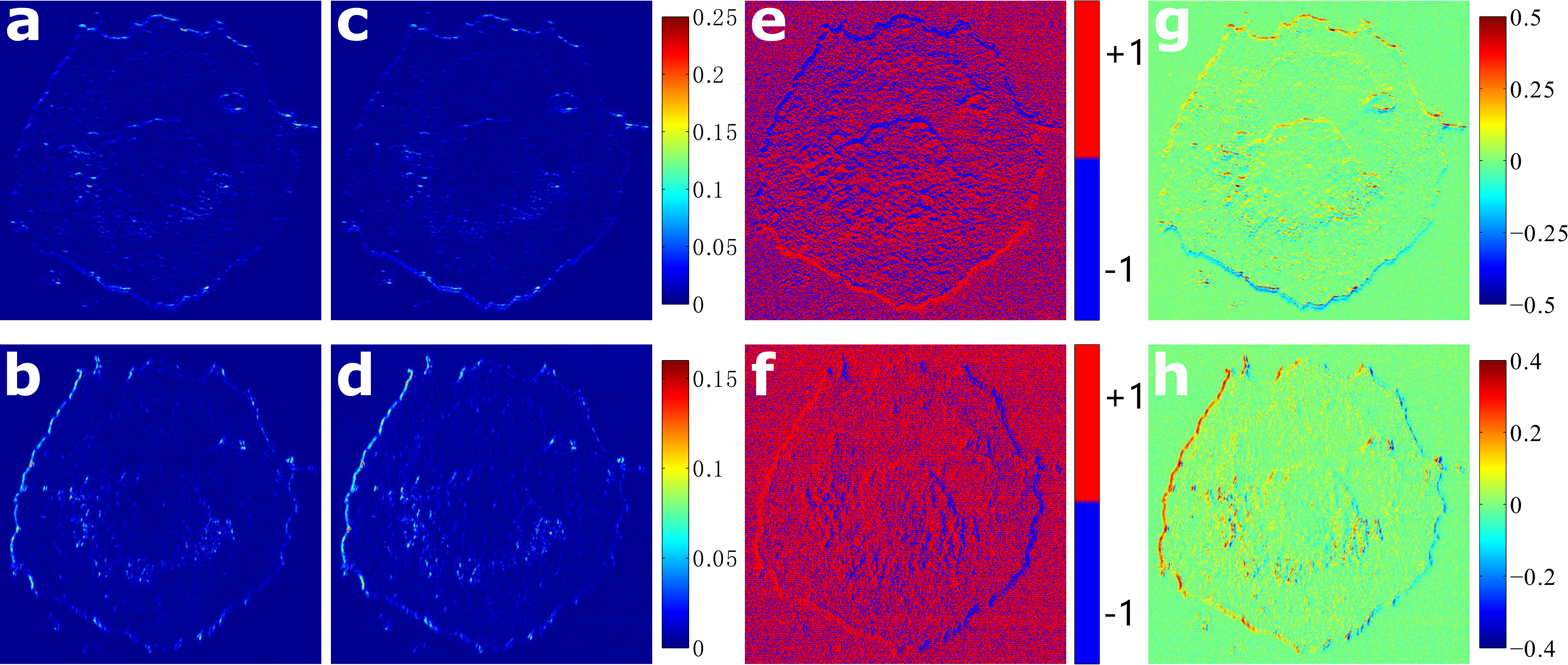}}%Here is how to import EPS art
\caption{\label{fig:S3} Experimental acquisition of directional derivatives of the phase distribution. (a, b) Non-biased differential contrast images with direction angles $\varphi  = 180^\circ $  and  $\varphi  = 90^\circ $, respectively. (c, d) Biased differential contrast images with a positive and very small bias as a perturbation on the basis of (a) and (bs), respectively. (e, f) Signs of the vertical and horizontal partial derivatives, respectively, consisting of the values of +1 and -1. (g, h) Acquired vertical and horizontal partial derivatives, respectively.}
\end{figure}

To acquire the directional derivatives of a phase distribution, we first experimentally measure the non-biased differential contrast images with  $b = 0$. For example, Figs.~\ref{fig:S3}(a, b) exhibit non-biased differential contrast images of the phase distribution [Fig.~4(b) in the main text] with direction angles $\varphi  = 180^\circ $  and  $\varphi  = 90^\circ $, respectively. We then introduce a positive and very small bias as a perturbation, by rotating the polarizer P2 to slightly increase the value of  ${\gamma _2}$. The resulted images are measured as Figs.~\ref{fig:S3}(c, d). Consequently, as we subtract Figs.~\ref{fig:S3}(a, b) from Figs.~\ref{fig:S3}(c, d), the signs of their difference values indicate the signs of the directional derivatives [Figs.~\ref{fig:S3}(e, f)], respectively.

Combining the obtained sign [Figs.~\ref{fig:S3}(e, f)] of  ${{\partial \psi } \mathord{\left/
 {\vphantom {{\partial \psi } {\partial l}}} \right.
 \kern-\nulldelimiterspace} {\partial l}}$ and the absolute values calculated from Figs.~\ref{fig:S3}(a, b), we can experimentally acquire the first-order directional derivatives of  $\psi \left( {x,y} \right)$. Figures~\ref{fig:S3}(g, h) show the obtained vertical and horizontal partial derivatives of phase distribution Fig.~4(b), respectively, which are the results shown as Figs.~5(a, b) in the main text.

%\bibliographystyle{osajnl}
%\bibliography{./References}

\end{document}